\documentclass[aps,preprintnumbers,amsmath,amssymb,11pt,floatfix]{revtex4}
\usepackage{graphicx}
\usepackage{subfigure}

\begin{document}

\title{Strain-induced band gaps in bilayer graphene}
\author{B. Verberck$^{1,2}$, B. Partoens$^{1}$, F.M. Peeters$^{1}$, and B. Trauzettel$^{2}$}
\affiliation{$^1$Departement Fysica, Universiteit Antwerpen, Groenenborgerlaan 171,
2020 Antwerpen, Belgium}
\affiliation{$^2$Institut f\"{u}r Theoretische Physik und Astrophysik,
Universit\"{a}t W\"{u}rzburg, Am Hubland, D-97070 W\"{u}rzburg, Germany}
\date{\today}

\begin{abstract}
We present a tight-binding investigation of strained bilayer graphene within linear elasticity theory, focusing on the different environments experienced by the A and B carbon atoms of the different sublattices.  We find that the inequivalence of the A and B atoms is enhanced by the application of perpendicular strain $\varepsilon_{zz}$, which provides a physical mechanism for opening a band gap, most effectively obtained when pulling the two graphene layers apart.  In addition, perpendicular strain introduces electron-hole asymmetry and can result in linear electronic dispersion near the K-point.
When applying lateral strain to one layer and keeping the other layer fixed, we find the opening of an indirect band gap for small deformations.
Our findings suggest experimental means for strain-engineered band gaps in bilayer graphene. 
\end{abstract}

\maketitle

\section{Introduction}
The synthesis of graphene --- a single layer of carbon atoms arranged into a
honeycomb structure --- and measurements of its electronic properties
 in 2004 by Novoselov {\it et al.} \cite{Nov1} has sparked off
the development of a whole new research field, continuing
to expand rapidly today in both experimental and theoretical directions.  The
high rate at which the graphene literature is growing is reflected by the
regular appearance of reviews on graphene in recent years; for a selection of
relevant accounts we refer to Refs.\
\cite{review1,Bee,review2,review3,review4,review5,Gei}.

The original excitement about graphene not only came from its
two-dimensionality \cite{Nov1,Nov2} --- flat two-dimensional (2D) crystals had
not been successfully fabricated before and were even predicted to be unstable
--- but also from its electronic properties \cite{Nov1,Nov2,Nov3,Kim}.  
Apart from novel fundamental physics, graphene boasts superior material
properties, including
high thermal conductivity, current density, carrier mobility,
carrier mean free path, strongness, stiffness, elasticity and
impermeability.

\begin{figure*}
\resizebox{8cm}{!}{\includegraphics{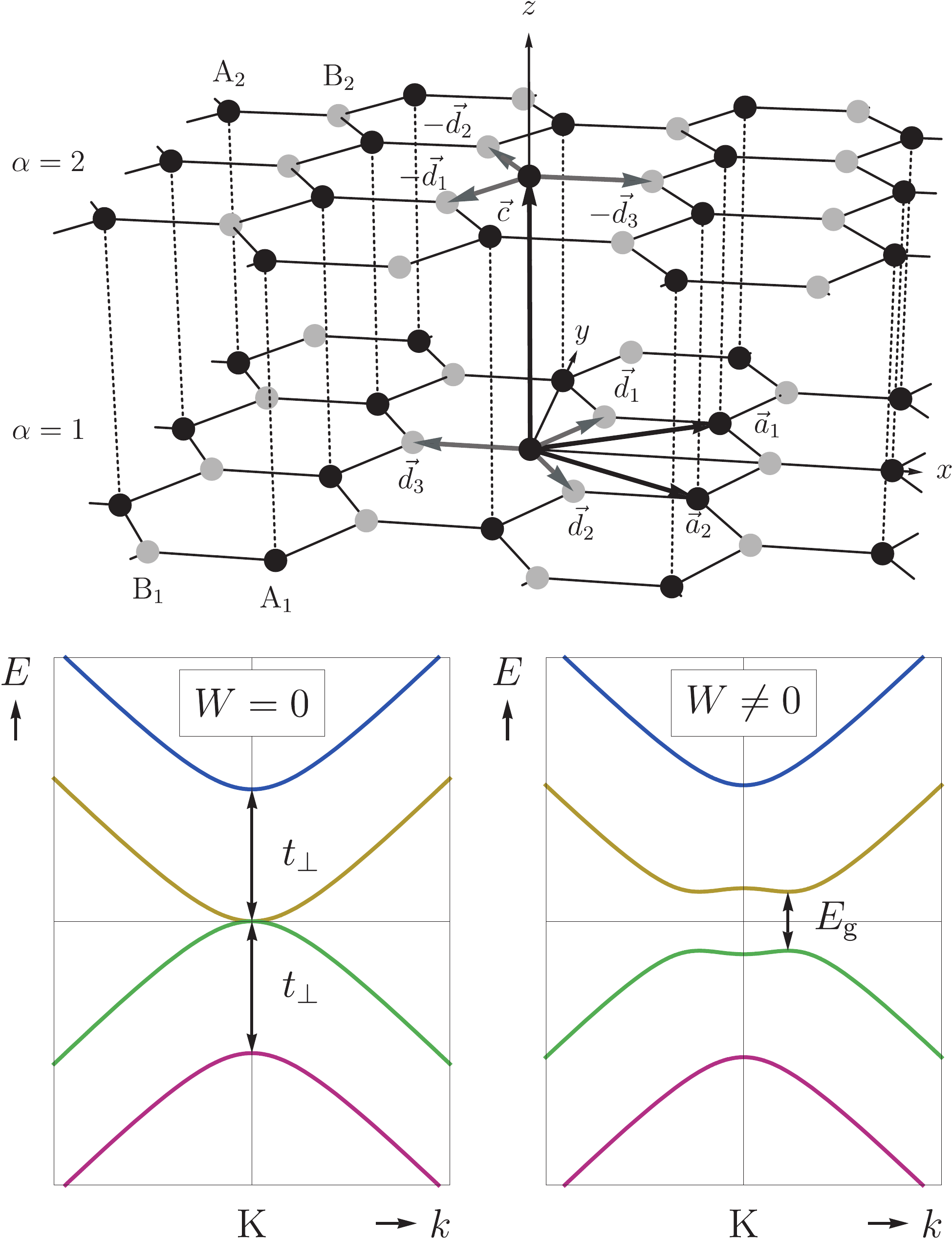}}
\caption{Graphene bilayer, with A carbon atoms in black and B carbon atoms in gray (top).  Electronic spectrum near the K-point in absence (bottom left) and presence (bottom right) of a bias voltage $W$ between the layers.
\label{fig1}}
\end{figure*}

Not only single graphene sheets but also stacks of a few
graphene layers \cite{Nov1} where the layers are coupled by van der Waals
interactions as in graphite can be isolated.  This has lead to investigations
of multilayer graphene, and in particular of bilayer graphene (Fig.\ \ref{fig1}, top).  Interestingly,
bilayer graphene displays (almost-)parabolic electronic dispersion at the
K-points (Fig.\ \ref{fig1}, bottom left), making electrons behave
differently
 \cite{Kos,Nil1,Par1,McC2} as compared to the single-layer case.
Bilayer graphene
offers the 
possibility of applying a bias voltage $W$ between the two layers, allowing to
tune the band structure.  In particular, the inequivalency of the two graphene
layers then gives rise to a ``Mexican-hat-like'' band structure featuring a band gap
\cite{McC2,Gui,Oht,McC3,Cas} of magnitude $E_\text{g} =
\frac{|eW|t_\perp}{\sqrt{(eW)^2 + t_\perp^2}}$ (Fig.\ \ref{fig1}, bottom right), with $t_\perp \approx 0.377$ eV
the interplane hopping parameter (see below).  A tunable energy gap is important for
possible electronic devices.

Another means of influencing (mono- or bilayer) graphene's electronic structure, currently receiving a lot of theoretical attention \cite{Gui2,Per, Lee, Mar,Muc,Muc2,Nan,Raz,Cho}, is provided
by mechanical deformations.  It has e.g.\ been shown that it is possible to
conceive inhomogeneous strains in single-layer graphene such that they act as a 
high uniform magnetic field, therefore resulting in strain-induced Landau
levels and a zero-field quantum Hall effect \cite{Gui2}.  
In Ref.\ \cite{Per}, Pereira {\it et al}.\ elaborated a tight-binding
description for uniaxially strained single-layer graphene and predicted that a
band gap can form upon deformations --- along preferred directions ---
beyond 20\%.
Recent investigations on strained bilayer graphene include
a standard tight-binding treatment of uniaxial strain \cite{Lee},
a description of elastic deformations and electron-phonon coupling in bilayer graphene by means of pseudo-magnetic gauge fields \cite{Mar}, the effect of strain on the Landau level spectrum and the quantum Hall effect
\cite{Muc,Muc2} and the effect of strain in combination with an external electric field \cite{Nan,Raz}. 

In the present paper, we employ a nearest-neighbor tight-binding description and linear elasticity theory to show that a perpendicular strain component modifies the on-site energies of the two carbon sublattices in bilayer graphene, which can open a band gap at the K-point.
The band gap is of a different nature than in the case of the ``Mexican-hat-like" electronic dispersions.
In addition, we consider the case of two graphene layers subjected to different (but uniform) strains, a scenario proposed recently and predicted --- by means of {\it ab initio} calculations --- to also lead to the opening of a band gap \cite{Cho}.  Uniform deformations parallel to the sheets are studied as well and are compared to the monolayer case \cite{Per}.

\section{Strained bilayer-graphene}\label{sec2}
We first consider the unstrained AB (Bernal) stacking variant of bilayer graphene: two graphene layers (labeled 1 and 2) at $c = 3.44$ {\AA}
apart with the A atoms of layer 2 sitting directly on top of the A atoms of
layer 1 (Fig.\ \ref{fig1}, top).  The B atoms of layer 1 and the B atoms of layer 2
have
no direct neighbor in the opposite layer.

The band structure of bilayer graphene can be described within the
tight-binding formalism.  The nearest-neighbor tight-binding hamiltonian
assumes one free $2p_z$ electron provided by each carbon atom and reads
\begin{multline}
   H  = V_\text{A} \sum_\alpha \sum_{\vec{X}_\alpha}  c_{\vec{X}_\alpha}^\dagger
c_{\vec{X}_\alpha} 
+ V_\text{B}\left( \sum_{\vec{X}_1}  c_{\vec{X}_1 + \vec{d}_1}^\dagger
c_{\vec{X}_1 + \vec{d}_1} + 
\sum_{\vec{X}_2}  c_{\vec{X}_2 - \vec{d}_1}^\dagger
c_{\vec{X}_2 - \vec{d}_1} \right)
\\
-t \Biggl( \sum_{\vec{X}_1} \sum_{l=1}^3   
\left[ c_{\vec{X}_1}^\dagger
c_{\vec{X}_1 + \vec{d}_l}
       + c_{\vec{X}_1 + \vec{d}_l}^\dagger c_{\vec{X}_1} \right]
 +
 \sum_{\vec{X}_2}  \sum_{l=1}^3   
\left[ c_{\vec{X}_2}^\dagger
c_{\vec{X}_2 - \vec{d}_l}
       + c_{\vec{X}_2 - \vec{d}_l}^\dagger c_{\vec{X}_2} \right] \Biggr) \\
 + t_\perp \sum_{\vec{X}_1}
  \left[c_{\vec{X}_1}^\dagger
c_{\vec{X}_1 + \vec{c}}
       + c_{\vec{X}_1 + \vec{c}}^\dagger c_{\vec{X}_1}\right].
\label{ham}
\end{multline}
The index $\alpha$ stands for the layer (1 or 2), the vectors $\vec{X}_\alpha$
are the lattice sites of the A atoms of layer $\alpha$.  Every A atom in layer 1 is
surrounded by three B atoms at relative position vectors $\vec{d}_1 =
\frac{1}{2}a\vec{e}_x + \frac{\sqrt{3}}{2}a \vec{e}_y$, $\vec{d}_2 =
\frac{1}{2}a\vec{e}_x - \frac{\sqrt{3}}{2}a \vec{e}_y$ and 
$\vec{d}_3  = -a\vec{e}_x$, where the bond length $a$ has
a value
of $1.42$ {\AA}, and the A atoms in layer 2 have neighboring B atoms at relative positions $-\vec{d}_l$, $l=1,2,3$.   The operators $c_{\vec{X}}^\dagger$ and 
$c_{\vec{X}}$ create and annihilate an electron at site
$\vec{X}$, respectively.
The
vector $\vec{c} = c\vec{e}_z$ ($c = 3.35$ {\AA}) connects two nearest-neighbor atoms in different
graphene sheets.
Values for the intra- and inter-plane hopping energies $t$ and $t_\perp$ are obtained from fitting the tight-binding model to experimental data for graphite.  Here we use the values quoted in Ref.\ \cite{Chu}: $t = 3.12$ eV and $t_\perp = 0.377$ eV.  The on-site energies $V_\text{A}$ and $V_\text{B}$ differ slightly due to the different environments of A and B atoms.   The difference $\Delta = |V_\text{A} - V_\text{B}| \approx 0.009$ eV \cite{Chu} is about two orders of magnitude smaller than the hopping parameter $t_\perp$ and is usually considered only in models going beyond the
nearest-neighbor tight-binding hamiltonian (\ref{ham}) where also A$_\text{1}$--B$_2$ and B$_1$--B$_2$ hoppings are taken into account (see e.g.\ Ref.\ \cite{Par1}).
We therefore put $V_\text{A}\approx V_\text{B} \equiv V$.  The value of the on-site energies $V$ will turn out to be of critical importance when considering perpendicular strain; we will return to it later.
For a comprehensive review on a complete tight-binding description of (unstrained) bilayer graphene, and in particular for a discussion of features in the electronic structure resulting from asymmetry of the diagonal (differences in on-site energies, $V_\text{A} \ne V_\text{B}$), we refer to Ref.\ \cite{Muc3}.

The direct-space hamiltonian (\ref{ham}) can be converted into a
reciprocal-space hamiltonian by introducing four-component spinors
\begin{subequations}
\begin{align}
\Psi^\dagger(\vec{q}) & = \left(\begin{array}{cccc}
c_{\text{A}_1}^\dagger(\vec{q}), &
c_{\text{B}_1}^\dagger(\vec{q}), &
c_{\text{A}_2}^\dagger(\vec{q}), &
c_{\text{B}_2}^\dagger(\vec{q}) \end{array}\right), \\
  \Psi(\vec{q}) & = \left( \begin{array}{c} c_{\text{A}_1}(\vec{q}) \\
c_{\text{B}_1}(\vec{q}) \\
c_{\text{A}_2}(\vec{q}) \\
c_{\text{B}_2}(\vec{q}) \end{array}\right), \\
H & =  \sum_{\vec{q}}\Psi^\dagger(\vec{q}) {\mathfrak H}(\vec{q})
\Psi(\vec{q}).
\end{align}
\end{subequations}
Here, the operators $c_i(\vec{q})$ ($i = \text{A}_1, \text{B}_1, \text{A}_2,
\text{B}_2$) are the discrete (lattice) Fourier transforms of $c_{\vec{X}_i}$:
\begin{align}
  c_i(\vec{q}) =
\frac{1}{\sqrt{N}}\sum_{\vec{X}_i}c_{\vec{X}_i}e^{-i\vec{q}\cdot \vec{X}_i},
\label{refsum}
\end{align}
where the $\vec{q}$-vectors of the first Brillouin zone are defined so that
the
properties $\sum_{\vec{q}} e^{i\vec{q}\cdot \vec{X}_i} = 
 N \delta_{\vec{X}_i,\vec{0}}$ and $\sum_{\vec{X}_i}
e^{i\vec{q}\cdot \vec{X}_i} = N \delta_{\vec{q},\vec{0}}$
hold.
The hamiltonian matrix $\mathfrak{H}(\vec{q})$ reads
\begin{align}
{\mathfrak H} (\vec{q}) 
 & = \left(\begin{array}{cccc}
 V  &  \zeta(\vec{q}) & t_\perp & 0 \\
 \zeta(\vec{q})^*  & V & 0 & 0 \\
 t_\perp & 0 & V &  \zeta(\vec{q})^* \\
 0 & 0 &  \zeta(\vec{q}) & V \\
  \end{array}\right), \label{mattot1}
\end{align}
with
\begin{align}
 \zeta(\vec{q}) = -t\sum_{l = 1}^3 e^{i\vec{q}\cdot \vec{d}_l}.
\end{align}
The band structure $\{E(\vec{q}) \}$ is obtained by solving the secular equation
$\det \bigl( \mathfrak{H}(\vec{q}) - E(\vec{q}) I_4 \bigr) = 0$, with $I_4$ the
$4\times 4$ unit matrix, in the first Brillouin zone.
At the K and K$^{'}$ points,
two electron energy bands touch each other at the Fermi
level, making the material a semi-metal (zero band gap).

Since the electronic properties of (bilayer) graphene are determined by 
the band structure near the K$^{(')}$ point --- $\vec{q}_\text{K} =
\frac{2\pi}{3 a}\vec{e}_x +  \frac{2\pi}{3 \sqrt{3} a} \vec{e}_y$
and 
$\vec{q}_{\text{K}^{'}} = \frac{2\pi}{3 a}\vec{e}_x - \frac{2\pi}{3 \sqrt{3} a}
\vec{e}_y$
  ---,
it is convenient to make a Taylor
expansion of $\mathfrak{H}(\vec{q})$ around $\vec{q}_{\text{K}^{(')}}$.  With
$\vec{q} = \vec{q}_{\text{K}} + \vec{k}$ and retaining only lowest-order
terms in $k$, the hamiltonian matrix near $\vec{q}_\text{K}$ becomes
\begin{align}
{\mathfrak H}^{\text{K}}(\vec{k}) 
 & = \left(\begin{array}{cccc}
 V  &  \frac{3at}{2} k e^{i\phi}e^{-i\frac{\pi}{6}} & t_\perp & 0 \\
 \frac{3at}{2} k e^{-i\phi}e^{i\frac{\pi}{6}}  & V & 0 & 0 \\
 t_\perp & 0 & V  &  \frac{3at}{2} k e^{-i\phi}e^{i\frac{\pi}{6}}   \\
 0 & 0 &  \frac{3at}{2} k e^{i\phi}e^{-i\frac{\pi}{6}}  & V  \\
  \end{array}\right). \label{mattot}
\end{align}
Here, $\phi$ is defined via $k_x + ik_y = k e^{i\phi}$.
For $\vec{q} = \vec{q}_{\text{K}^{'}} + \vec{k}$,
the hamiltonian matrix ${\mathfrak H}^{\text{K}^{'}}(\vec{k})$ is the complex
conjugate of ${\mathfrak H}^{\text{K}}(\vec{k})$.
Note that the quantity $\frac{3at}{2}$ can be rewritten as $\hbar
v_\text{F}$, with $v_\text{F}$ the Fermi velocity.  Diagonalisation of (\ref{mattot}) results in the dispersion shown in the left bottom of Fig.\ \ref{fig1}.

As mentioned in the introduction, the application of a bias voltage $W$ between layers 1 and 2
(replacing the elements ${\mathfrak
H}^{\text{K}^{(')}}_{11}(\vec{k})$ and
${\mathfrak H}^{\text{K}^{(')}}_{33}(\vec{k})$ by $eW/2$ and
${\mathfrak
H}^{\text{K}^{(')}}_{22}(\vec{k})$ and
${\mathfrak H}^{\text{K}^{(')}}_{44}(\vec{k})$ by $-eW/2$) results in a finite band gap \cite{McC2,Gui,Oht,McC3,Cas}.
However, this band gap lies not at $\vec{q} =
\vec{q}_{\text{K}^{(')}}$ but at a $q$-vector slightly away from
$q_{\text{K}^{(')}}$ (Fig.\ \ref{fig1}, bottom right), and its theoretical maximum value is $\lim_{W \longrightarrow \infty} E_\text{g} = t_\perp$.
In the following, we will show that applying a strain to the bilayer
graphene lattice can also lead to a band gap, which is not bounded by
$t_\perp$.

In the presence of a displacement field $\vec{u}(\vec{X})$,
the position of an atom formerly at
$\vec{X}$ is  $\vec{X} + \vec{u}(\vec{X})$.
The effect of a small deformation of the lattice can be written as a
correction $\delta H$ to the original hamiltonian $H$, with
\begin{multline}
   \delta H  =\delta V_\text{A} \sum_\alpha \sum_{\vec{X}_\alpha}  c_{\vec{X}_\alpha}^\dagger
c_{\vec{X}_\alpha} 
+ \delta V_\text{B}\left( \sum_{\vec{X}_1}  c_{\vec{X}_1 + \vec{d}_1}^\dagger
c_{\vec{X}_1 + \vec{d}_1} + 
\sum_{\vec{X}_2}  c_{\vec{X}_2 - \vec{d}_1}^\dagger
c_{\vec{X}_2 - \vec{d}_1} \right) \\
- \sum_{\vec{X}_1}\sum_{l=1}^3 \delta t_l
\left[ c_{\vec{X}_1}^\dagger
c_{\vec{X}_1 + \vec{d}_l}
       + c_{\vec{X}_1 + \vec{d}_l}^\dagger c_{\vec{X}_1} \right]
- \sum_{\vec{X}_2}\sum_{l=1}^3 \delta t_l
\left[ c_{\vec{X}_2}^\dagger
c_{\vec{X}_2 - \vec{d}_l}
       + c_{\vec{X}_2 - \vec{d}_l}^\dagger c_{\vec{X}_2} \right] \\
 + \sum_{\vec{X}_1} \delta t_\perp 
  \left[c_{\vec{X}_1}^\dagger
c_{\vec{X}_1 + \vec{d}}
       + c_{\vec{X}_1 + \vec{d}}^\dagger c_{\vec{X}_1}\right].
\label{hamcorr}
\end{multline}
Within the nearest-neighbor tight-binding approximation, the changes in on-site
and hopping energy parameters $V_\text{A}$, $V_\text{B}$, $t$ and $t_\perp$
 are
related to changes in nearest-neighbor interatomic distances (bond lengths). 
We stress that it is therefore important to distinguish between A and B sites since they have
different environments in the bilayer.
As pointed out before, an A site has three
in-plane nearest-neighbor B sites and one neighboring A site in the opposite
layer at a distance $c + \delta c$; a B site has only the three surrounding
in-plane A sites as nearest neighbors (see Fig.\ \ref{fig1}, top). Denoting the
three (not
necessarily equal) changed bond lengths between neighboring in-plane A and B
atoms by $a_l = a + \delta a_l$ ($l = 1,2,3$), we have for the
corrections to the on-site energies to linear order in the deformations
\begin{subequations}
\begin{align}
 \delta V_\text{A} & =  \sum_{l = 1}^3\left.\frac{\partial V}{\partial
a }\right|_a \delta a_l
                   + \left.\frac{\partial V}{\partial c}\right|_c \delta c,
\label{bc1} \\
 \delta V_\text{B} & = \sum_{l = 1}^3\left.\frac{\partial V}{\partial
a }\right|_a \delta a_l
\end{align}
\end{subequations}
for each of the two layers.  For the hopping parameters we have
\begin{subequations}
\begin{align}
 \delta t_l & = \left.\frac{\partial t}{\partial
a}\right|_a \delta a_l, \\
  \delta t_\perp & = \left.\frac{\partial t_\perp}{\partial
c}\right|_c \delta c. \label{bc4}
\end{align}
\end{subequations}
In the following we will drop the attributes $|_a$ and $|_c$.
The link between the corrections $\delta V_\text{A}$, $\delta V_\text{B}$,
$\delta t_l$ and $\delta t_\perp$ and the displacement field $\vec{u}(\vec{X})$
then comes from considering the bond length corrections $\delta a$ and $\delta
c$.  Details of the calculations and approximations involved are given in
Appendix \ref{appA}; the resulting correction $\delta {\mathfrak H}(\vec{q})$ to the hamiltonian matrix ${\mathfrak
H}(\vec{q}) $ reads

{\scriptsize
\begin{multline}
\delta{\mathfrak H}(\vec{q}) =    \left(\begin{array}{cccc}
 \frac{3a}{2}\frac{\partial V}{\partial a}(\varepsilon_{xx} + \varepsilon_{yy})
 + c\frac{\partial V}{\partial c}\varepsilon_{zz} & \delta \zeta(\vec{q})
 & c \frac{\partial t_\perp}{\partial c} \varepsilon_{zz}  & 0 \\
 \delta \zeta(\vec{q})^* & \frac{3a}{2}\frac{\partial V}{\partial
a}(\varepsilon_{xx} + \varepsilon_{yy}) 
 & 0 & 0 \\
 c \frac{\partial t_\perp}{\partial c} \varepsilon_{zz} & 0 &
\frac{3a}{2}\frac{\partial V}{\partial a}(\varepsilon_{xx} + \varepsilon_{yy})
 + c\frac{\partial V}{\partial c}\varepsilon_{zz}  &   \delta \zeta(\vec{q})^* \\
 0 & 0 &  \delta \zeta(\vec{q}) &
\frac{3a}{2}\frac{\partial V}{\partial a}(\varepsilon_{xx} + \varepsilon_{yy})
  \\
  \end{array}\right), \label{corr1}
\end{multline}
}
with
\begin{align}
  \delta \zeta(\vec{q}) & = -\sum_{l = 1}^3 \delta t_l e^{i \vec{q}\cdot \vec{d}_l} \nonumber \\
   & =
   - a \frac{\partial t}{\partial a} e^{i \frac{1}{2}a q_x} e^{i\frac{\pi}{6}} \Big(\bigl[ \frac{1}{2}   \cos (\frac{\sqrt{3}}{2}a q_y)  + e^{-i  \frac{3}{2}a q_x}\bigr] \varepsilon_{xx}   + \frac{3}{2}   \cos (\frac{\sqrt{3}}{2}a q_y)  \varepsilon_{yy} 
   + \sqrt{3} i   \sin (\frac{\sqrt{3}}{2}a q_y) \varepsilon_{xy}
    \Bigr). \label{corr1with}
\end{align}
Here, the quantities $\varepsilon_{ij}$ ($i,j = x, y, z$)
are elements of the strain tensor $[\varepsilon]$, the general definition of
which involve derivatives of the displacement field components to the
coordinates.  For the uniform displacements we mostly consider in this work the elements
$\varepsilon_{ij}$ can be related to relative increments/decrements of the bond
lengths occuring in the lattice (see Appendix \ref{appA}): $\vec{u}(\vec{X}) =[\varepsilon] \vec{X}$.

The leading correction $\delta {\mathfrak
H}^{\text{K}} $ to the hamiltonian matrix ${\mathfrak
H}^{\text{K}} (\vec{k}) $ at the K-point is $k$-independent and reads

{\scriptsize
\begin{multline}
\delta{\mathfrak H}^{\text{K}} =    \left(\begin{array}{cccc}
 \frac{3a}{2}\frac{\partial V}{\partial a}(\varepsilon_{xx} + \varepsilon_{yy})
 + c\frac{\partial V}{\partial c}\varepsilon_{zz} & -i
\frac{3a}{4}\frac{\partial t}{\partial a} (-\varepsilon_{xx} +
\varepsilon_{yy} +
2i\varepsilon_{xy})
 & c \frac{\partial t_\perp}{\partial c} \varepsilon_{zz}  & 0 \\
 i \frac{3a}{4}\frac{\partial t}{\partial a} (-\varepsilon_{xx} + 
\varepsilon_{yy} 
-2i\varepsilon_{xy})  & \frac{3a}{2}\frac{\partial V}{\partial
a}(\varepsilon_{xx} + \varepsilon_{yy}) 
 & 0 & 0 \\
 c \frac{\partial t_\perp}{\partial c} \varepsilon_{zz} & 0 &
\frac{3a}{2}\frac{\partial V}{\partial a}(\varepsilon_{xx} + \varepsilon_{yy})
 + c\frac{\partial V}{\partial c}\varepsilon_{zz}  &  i
\frac{3a}{4}\frac{\partial t}{\partial a} (-\varepsilon_{xx} + 
\varepsilon_{yy} -
2i\varepsilon_{xy}) \\
 0 & 0 & - i \frac{3a}{4}\frac{\partial t}{\partial a} (-\varepsilon_{xx} + 
\varepsilon_{yy} +
2i\varepsilon_{xy}) &
\frac{3a}{2}\frac{\partial V}{\partial a}(\varepsilon_{xx} + \varepsilon_{yy})
  \\
  \end{array}\right). \\ \label{corr}
\end{multline}
}

\hspace*{-\parindent}For $\vec{q} = \vec{q}_{\text{K}^{'}} + \vec{k}$,
the hamiltonian matrix correction $\delta{\mathfrak H}^{\text{K}^{'}}$
is the complex
conjugate of $\delta {\mathfrak H}^{\text{K}}$.
The K-point hamiltonian correction for a strained single graphene layer, of the form
\begin{align}
  \left(\begin{array}{cc} g_1 (\varepsilon_{xx} + \varepsilon_{yy}) & g_2 (\varepsilon_{xx} - \varepsilon_{yy} + 2i\varepsilon_{xy})  \\
  g_2^* (\varepsilon_{xx} - \varepsilon_{yy} - 2i\varepsilon_{xy}) & g_1 (\varepsilon_{xx} + \varepsilon_{yy}  )
  \end{array}\right), \label{hammono}
\end{align}
has been derived before in the context of electron-phonon coupling in carbon nanotubes \cite{Suz}, where the term proportional to $g_1$ is a
deformation potential, a concept going back to Bardeen and Shockley \cite{Bar} and the term proportional to $g_2$ corresponds to a bond-length change.  However, to our knowledge, the derivation of the strained bilayer hamiltonian [Eqs.\ (\ref{corr1}) -- (\ref{corr1with})] as given in Appendix \ref{appA} --- with particular emphasis on the asymmetry on the diagonal --- has not been reported before.

\section{Perpendicular uniform strain}\label{sec3}
We first consider the bilayer-specific possibility of perpendicular strain,
$\varepsilon_{zz}$, associated with a change from the inter-layer distance $c$ 
to $c' = c(1 + \varepsilon_{zz})$.  Putting the
in-plane strain tensor elements
to zero, the hamiltonian ${\mathfrak H}^\text{K}(\vec{k}) + \delta{\mathfrak
H}^\text{K}$ takes on the form
\begin{align}
  {\mathfrak H}^\text{K}(\vec{k}) + \delta{\mathfrak H}^\text{K} =
\left(\begin{array}{cccc}
             V + c\frac{\partial V}{\partial c}\varepsilon_{zz} & \hbar
v_\text{F} ke^{i\phi} & t_\perp + c\frac{\partial t_\perp}{\partial
c}\varepsilon_{zz} & 0 \\
             \hbar v_\text{F} ke^{-i\phi} & V & 0       & 0 \\
             t_\perp + c\frac{\partial t_\perp}{\partial
c}\varepsilon_{zz} & 0 & V + c\frac{\partial V}{\partial c}\varepsilon_{zz} &
\hbar v_\text{F} ke^{-i\phi} \\
             0 &       0 & \hbar v_\text{F} ke^{i\phi} & V \\
            \end{array}\right). \label{refH}
\end{align}
The phase factors $e^{\mp i\frac{\pi}{6}}$ present in 
Eq.\ (\ref{mattot}) have been eliminated by means of a redefinition of the
operators for B sites [Eqs.\ (\ref{redef1}) -- (\ref{redef2})].

\begin{figure*}
\resizebox{16cm}{!}
{\includegraphics{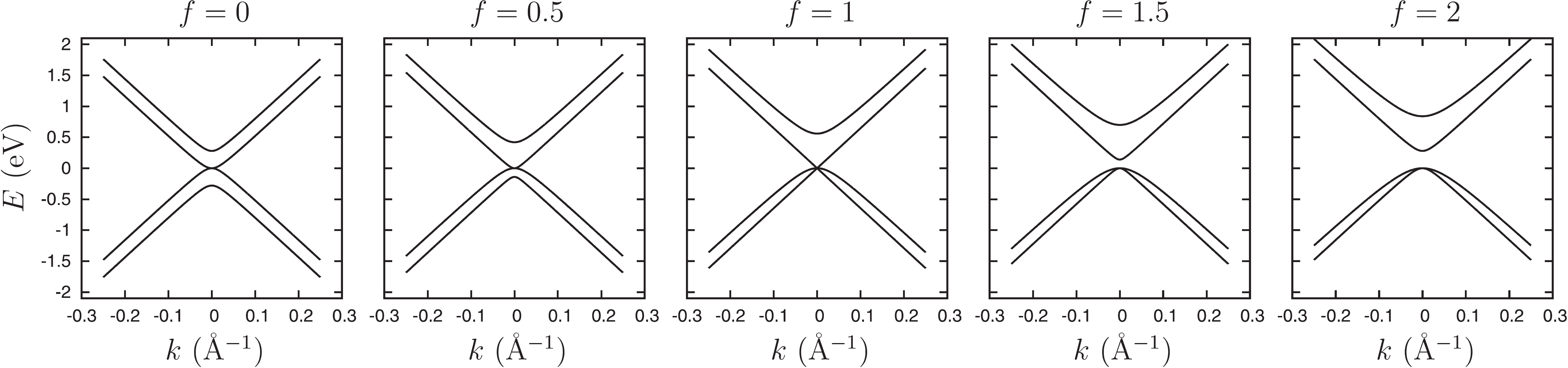}}
\caption{Low-$k$ band structure $E(k)$ for perpendicularly uniformly strained bilayer graphene for various values of
$\left| c\frac{\partial V}{\partial c} \varepsilon_{zz} \right| = f \left(
t_\perp + c\frac{\partial t_\perp}{\partial c} \varepsilon_{zz} \right)$.
\label{fig2}}
\end{figure*}

The symmetry-breaking along the diagonal exhibited by the hamiltionian matrix
(\ref{refH}) is unusual since it distinguishes not
between layers (as e.g.\ a bias voltage between the two layers would do) but
between A and B sublattices.  Interestingly, the hamiltonian (\ref{refH}) displays
the possibility of the opening of a band gap at $k = 0$: if
\begin{align}
  \left| c\frac{\partial V}{\partial c} \varepsilon_{zz} \right| >
t_\perp + c\frac{\partial t_\perp}{\partial c} \varepsilon_{zz} , \label{crit}
\end{align}
the degeneracy of the bands at $k = 0$ is lifted, as illustrated in Fig.\
\ref{fig2}.  (Criterion (\ref{crit}), an analytical result, holds when $t_\perp + c\frac{\partial
t_\perp}{\partial c} \varepsilon_{zz} > 0$, i.e.\ when $c\left|\frac{\partial
t_\perp}{\partial c} \varepsilon_{zz}\right|$ is small compared to $t_\perp$.)
Recently, Mucha-Kruczy\'{n}ski et al.\ \cite{Muc3} examined the asymmetry of the (unstrained) graphene bilayer hamiltonian's diagonal.  The possibility of band gap openings and electron-hole asymmetry, as encountered here in Fig.\ \ref{fig2}, due to different on-site energies, was realized.  Here, we show that strain enhances the diagonal's asymmetry [Eq.\ (\ref{corr1})] and that elastic deformations therefore provide a physical mechanism for the band structure modifications discussed in Ref.\ \cite{Muc3}.

\begin{figure}
\resizebox{8cm}{!}
{\includegraphics{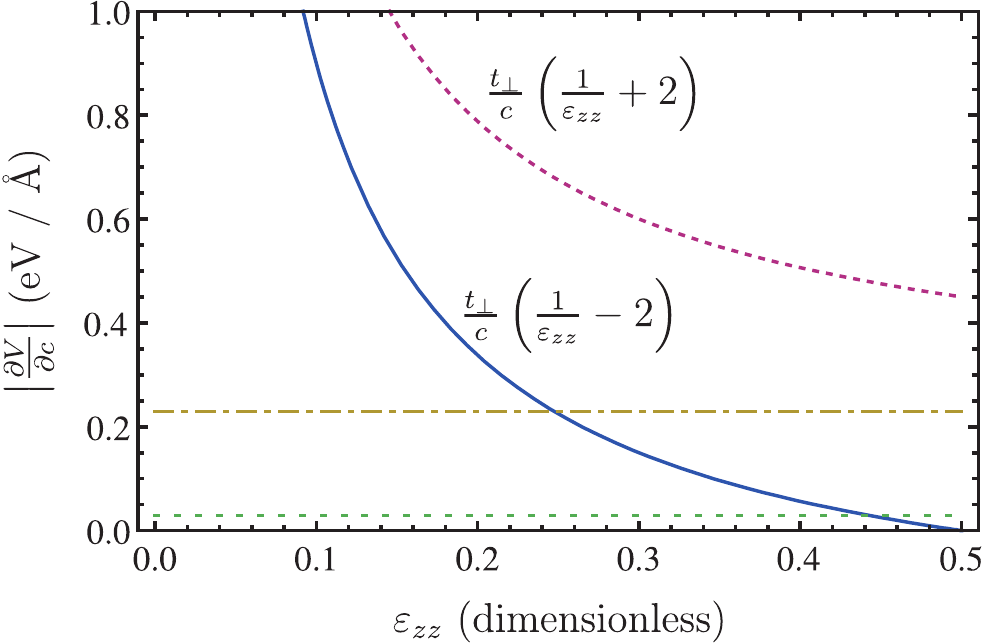}}
\caption{(Color online)  Plots of
the variation of $\frac{t_\perp}{c}\left(\frac{1}{\varepsilon_{zz}}-2\right)$ 
(blue, full line, expansion) and $\frac{t_\perp}{c}\left(\frac{1}{ \varepsilon_{zz} }+2\right)$
(violet, dashed line, contraction) with $\varepsilon_{zz}$.  The two horizontal lines represent the values of $\left| \frac{\partial V}{\partial c} \right|$ resulting from using the two extreme values for $V$ considered in the present work --- $V = -2$ eV (olive, dashed-dotted line) and $V = -0.25$ eV (green, dashed line).
\label{fig3}}
\end{figure}

To check whether the possibility of a strain-induced band gap is
experimentally relevant, a more quantitative investigation of criterion (\ref{crit}) is
in order. First, the variation of the hopping parameter $t_\perp$ with
interlayer distance $c$ can be estimated using Harrison's relation
\begin{align}
  \frac{\partial t_\perp}{\partial c} \approx -2\frac{t_\perp}{c}, \label{relchange}
\end{align}
which follows from an assumed inverse-square dependence of $t_\perp$ on $c$ \cite{Har}.
The inequality (\ref{crit}) then becomes
\begin{align}
  c \left| \frac{\partial V}{\partial c}\right| \left| \varepsilon_{zz} \right|
\gtrsim t_\perp(1 - 2\varepsilon_{zz}).
\end{align}
For expansion ($\varepsilon_{zz} > 0$), this can be rewritten as
\begin{align}
  c \left| \frac{\partial V}{\partial c}\right|  
    \gtrsim t_\perp\left(\frac{1}{\varepsilon_{zz}} - 2\right),
\end{align}
while for contraction ($\varepsilon_{zz} < 0$), one obtains
\begin{align}
  c \left| \frac{\partial V}{\partial c}\right|  
    \gtrsim t_\perp\left(-\frac{1}{\varepsilon_{zz}} + 2\right).
\end{align}
The intra- and inter-plane hopping parameters have values of $t = 3.12$ eV and
$t_\perp = 0.377$ eV, respectively.
The interplane
distance is $c = 3.35$ {\AA}, so that $\frac{t_\perp}{c} = 0.113$ eV/{\AA}. 
Using these values, we have plotted the quantities
$\frac{t_\perp}{c}\left(\frac{1}{\varepsilon_{zz}}-2\right)$ 
and $\frac{t_\perp}{c}\left(\frac{1}{|\varepsilon_{zz}|}+2\right)$
in Fig.\ \ref{fig3}.  Interestingly, it follows that the
band gap formation criterion is reached more easily (smaller
$|\varepsilon_{zz}|$) in the case of expansion.

\begin{figure}
\resizebox{8cm}{!}
{\includegraphics{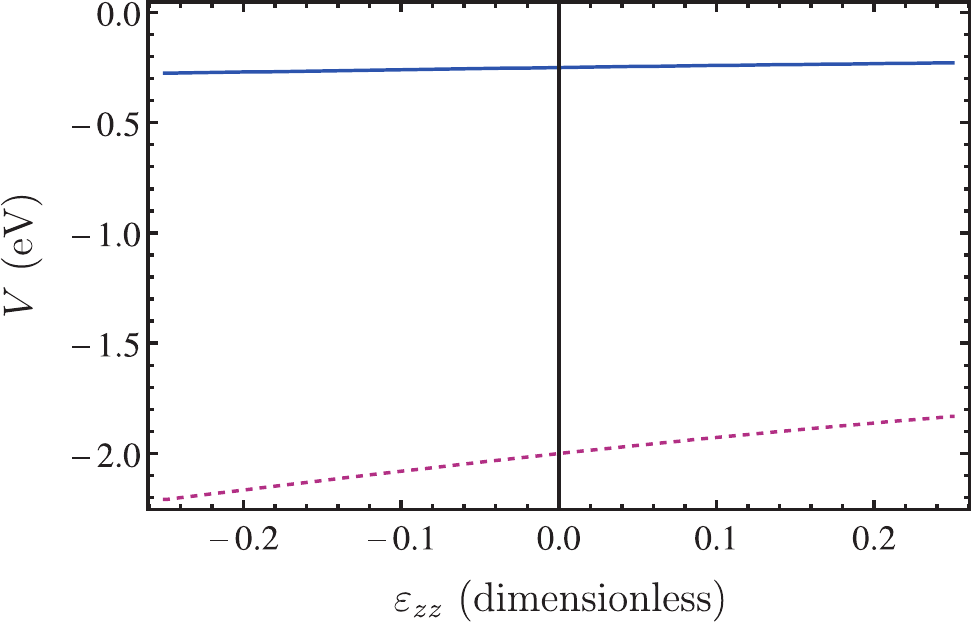}}
\caption{Plots of
the variation of $V(c')$ with $c' = c(1 + \varepsilon_{zz})$ for $V(c) = -0.25$ eV 
(blue, full line) and $V(c) = -2$ eV (violet, dashed line).
\label{fig4}}
\end{figure}

To obtain a physically meaningful estimate for the quantity $\left|
\frac{\partial V}{\partial c} \right|$ we proceed as
follows. First, we recall the tight-binding definition of $V$:
\begin{align}
  V \equiv \varepsilon_{2p_z} = \int d\vec{r} \psi^*(\vec{r})
U(\vec{r})\psi(\vec{r}). \label{definition}
\end{align}
Here, $\psi(\vec{r})$ is the $2p_z$ electron wave function for a carbon atom,
and $U(\vec{r})$ is the periodic potential of the lattice.
Suprisingly, while the tight-binding formalism is a standard method for
modelling band structures --- especially for carbon structures ---, a numerical
value for $V \equiv \varepsilon_{2p_z}$ has not yet been established in the
literature.  The plausible reason is that when the difference between $V_\text{A}$ and $V_\text{B}$ is neglected (see Sect.\ \ref{sec2}), the presence of $V$ on the diagonal of the hamiltonian
matrix merely results in an overall shift of the energy bands, making its actual value redundant.  In Ref.\
\cite{Rei}, Reich {\it et al.} present a detailed comparison of tight-binding
and ab initio energy dispersion calculations for graphene and provide fits of
the tight-binding parameters to the band structures obtained by
density-functional theory.  Two variants were considered: one fitting the
full $M\Gamma K M$-path in $k$-space and one where only $\vec{k}$-vectors
yielding optical transitions with energies less than 4 eV were taken into
account, resulting in $\varepsilon_{2p_z} = -0.28$ eV and $-2.03$ eV,
respectively.  We will therefore consider the range $-0.25$ eV $
\ge \varepsilon_{2p_z} \ge -2$ eV.
The simplest way of modelling $U(\vec{r})$ is to put positively charged ions
(charge $1e>0$) at $\vec{r} = \vec{0}$ and $\vec{r} = c'\vec{e}_z$:
\begin{align}
   U(\vec{r}) \propto -\frac{1}{|\vec{r}|} - \frac{1}{|\vec{r} - c'\vec{e}_z|} =
-\frac{1}{r} - \frac{1}{\sqrt{r^2 - 2c'r\cos\theta + c'^2}}, \label{propto1}
\end{align}
where spherical coordinates have been introduced.
For the carbon $2p_z$ electron wave function, we follow the common practice of
taking the hydrogen-like $2p_z$ orbital:
\begin{align}
 \psi(\vec{r}) \propto r e^{-\frac{r}{2a_0}}\cos\theta, \label{propto2}
\end{align}
where $a_0 = 0.529$ {\AA} is the Bohr radius.  We then get for $V$ the
expression
\begin{align}
  V = C \int_0^\infty r^2 dr \int_0^\pi \sin\theta d\theta r^2
e^{-\frac{r}{a_0}} \cos^2 \theta\left( -\frac{1}{r} - \frac{1}{\sqrt{r^2 -
2c'r\cos\theta + c'^2}}\right), \label{Uexpr}
\end{align}
where the proportionality factors left unspecified in Eqs.\ (\ref{propto1}) and
(\ref{propto2}), together with the factor $2\pi$ coming from the azimuthal
integration, have been collected into the factor $C$.  The integrals 
\begin{align}
 S_{nc'}(r) = \int_0^\pi d\theta \frac{\sin\theta \cos^2\theta}{\sqrt{r^2 - 2nc'
r \cos \theta + (nc')^2}}
\end{align}
entering Eq.\ (\ref{Uexpr}) can be
solved analytically:
\begin{subequations}
\begin{align}
   S_0(r) & = \frac{2}{3 r}, \\
   S_{nc' > 0}(r) & = \left\{ \begin{array}{l} \frac{2}{15} \frac{5(nc')^2 +
2r^2}{(nc')^3}\text{ if $0\le r \le nc'$} \\
  \frac{2}{15}\frac{2(nc')^2 + 5r^2}{r^3}\text{ if $r > nc'$}
 \end{array}\right. .
\end{align}
\end{subequations}
For $c' = c = 3.35$ {\AA},
the integral
\begin{align}
 D(c) = \int_0^\infty dr r^4 e^{-\frac{r}{a_0}}\bigl(S_0(r) + S_{1c}(r)\bigr)
\end{align}
has the value $D(c) = 0.539$ {\AA}$^4$.  From $V = C D(c)$ we can then
obtain the ``calibration value'' $C = V/D(c)$.  For the
values $V = -0.25$ eV and $V = -2$ eV, we obtain $C = -0.463$ eV/{\AA}$^{-4}$
and $C = -3.709$ eV/{\AA}$^{-4}$, respectively.
The dependence of $V$ on $c'$ reads $V(c') = C D(c')$, it is visualised in
Fig.\ \ref{fig4}.  Clearly,
a linear approximation in the range $-0.25 \le \varepsilon_{zz} \le 0.25$ is
justifiable, and the value of $\frac{\partial V}{\partial c}
$
can be easily determined
numerically.  We find
$ \frac{\partial V}{\partial c} = 0.0287$ eV/{\AA}  and $0.230$ eV/{\AA}  for
$V = -0.25$ eV and $V = -2$ eV, respectively; these values are marked by
horizontal lines in Fig.\ \ref{fig3}.  

Within the present model, it follows that in the case of a large
tight-binding on-site energy parameter ($|V| = 2$ eV), a band gap opens 
for positive strains larger than $\varepsilon_{zz}\approx 0.25$ (see Fig.\
\ref{fig3}), corresponding to interlayer distances larger than $c' \approx 4.19$
{\AA}.  For $\varepsilon_{zz}\approx 30\%$, the band gap's magnitude is about $125$ meV (Fig.\ \ref{fig2}, $f = 1.5$).

We recall that formally, criterion (\ref{crit}) is valid when the right-hand side is positive, i.e.\ when $\varepsilon_{zz} < 0.5$, hence the upper limit of $\varepsilon_{zz} = 0.5$ in Fig.\ \ref{fig3}.  Going beyond strains of $50\%$ (both positive and negative) would be well outside the validity of the assumption of small displacements $\vec{u}(\vec{X})$, on which the hamiltonian matrix (\ref{corr1}) relies (see Appendix \ref{appA}).  For $\varepsilon \sim 0.25$, neglected contributions are of the order $\varepsilon^2 \sim 0.063$ which is still acceptable.  Similarly, we point out that the unphysical behavior of an ever increasing band gap with increasing $\varepsilon_{zz}$ must become invalid when linear elasticity, i.e.\ the assumption of small bond length changes [Eq.\ (\ref{A2})], fails.   

Based on the
foregoing elaborations, stating that the opening
of a strain-induced band gap by pulling apart the two graphene layers may be
experimentally observed is a fair conclusion.  Our main purpose here is not to
provide accurate predictions but rather to point out the consequences of the
symmetry-breaking along the diagonal in the hamiltonian of strained bilayer
graphene.  Accurate density-functional theory calculations could provide better
numerical estimates, but most importantly, experiments should be undertaken to
investigate the effect on the band structure upon pushing together or pulling
apart the graphene layers.


\section{Symmetric uniform $xy$-strain}\label{longitudinal}
We next consider pulling or pushing the bilayer in a direction parallel to the graphene sheets ($\varepsilon_{zz} = 0$).  The tight-binding hamiltonian then reads

{\scriptsize
\begin{multline}
{\mathfrak H}(\vec{k}) =    \left(\begin{array}{cccc}
 V + \frac{3a}{2}\frac{\partial V}{\partial a} (\varepsilon_{xx} + \varepsilon_{yy}) & \zeta(\vec{q}) + \delta \zeta (\vec{q})  & t_\perp & 0 \\
 \zeta^*(\vec{q}) + \delta \zeta^* (\vec{q}) & V + \frac{3a}{2}\frac{\partial V}{\partial a} (\varepsilon_{xx} + \varepsilon_{yy}) & 0 & 0 \\
 t_\perp & 0 &
V + \frac{3a}{2}\frac{\partial V}{\partial a} (\varepsilon_{xx} + \varepsilon_{yy})
  &   \zeta^*(\vec{q}) + \delta \zeta^* (\vec{q}) \\
 0 & 0 &  \zeta(\vec{q}) + \delta \zeta (\vec{q}) & V + \frac{3a}{2}\frac{\partial V}{\partial a} (\varepsilon_{xx} + \varepsilon_{yy})
  \\
  \end{array}\right). \\
\end{multline}
}
The $4$ solutions of the corresponding secular equation are
\begin{align}
  E_{\pm,\pm}(\vec{q}) = \pm \sqrt{\frac{t_\perp^2 + 2|\zeta(\vec{q}) + \delta \zeta(\vec{q})|^2 
       \pm t_\perp \sqrt{t_\perp^2 + 4|\zeta(\vec{q}) + \delta \zeta(\vec{q})|^2}}{2}}. \label{secxy}
\end{align}
There can only be a band gap when $\zeta(\vec{q}) + \delta \zeta(\vec{q})$ differs from zero for all $\vec{q}$-vectors.  The same condition arises from the monolayer strain problem, where the $2$ solutions of the secular equation read
$E_\pm(\vec{q}) = \pm| \zeta(\vec{q}) + \delta \zeta(\vec{q}) |$ (leading to the Dirac cone at the K$^\text{($'$)}$-point for $\delta \zeta(\vec{q}) = 0$).  Pereira {\it et al.} \cite{Per} have investigated the behavior of $| \zeta(\vec{q}) + \delta \zeta(\vec{q}) |$ in detail.  Using the isotropy of a 2D hexagonal lattice, the strain tensor can be written as
\begin{align}
[\varepsilon] = \left( \begin{array}{cc}
  \cos^2 \theta - \sigma \sin^2 \theta & (1 + \sigma) \cos\theta \sin\theta \\
  (1 + \sigma) \cos\theta \sin\theta & \sin^2 \theta - \sigma \cos^2 \theta 
  \end{array} \right) \label{isotropic},
\end{align}
with $\theta$ the angle between the tension $\vec{T}$ and the $y$-axis ($\vec{T} = T\cos(\pi/2 + \theta)\vec{e}_x + T\sin(\pi/2 + \theta)\vec{e}_y$) \cite{Perfootnote}, and $\sigma$ the Poisson ratio, which takes the graphite value of $0.165$ which we choose in the remainder \cite{Bla}.  The conclusions made by Pereira {\it et al.} are that (i) the minimal strain required for opening a gap is about $23\%$, that (ii) tension along the zig-zag direction ($\theta = 0, \pi/3, 2\pi/3$) is optimal for the opening of a band gap, and that (iii) tension along the armchair direction ($\theta = \pi/2, 5\pi/6, 11\pi/6)$ never results in a band gap.
From Eq.\ (\ref{secxy}) it follows that the inter-plane coupling $t_\perp$ does not play any role and that the same conclusions are valid for the bilayer.

\section{Asymmetric uniform $xy$-strain}
Finally, we consider the possibility of applying different strains (parallel to
the bilayer) to the two layers $\alpha = 1$ and $\alpha = 2$.  In Ref.\
\cite{Cho}, ab initio calculations of a graphene bilayer's band structure were
performed for the situation where one of the two layers is subjected to positive
strain along the zig-zag or armchair direction (without the elastic response in
the perpendicular direction).  Here, we consider a more general situation.
We choose to let layer $1$ intact ($[\varepsilon] = [0]$), while layer $2$
is pushed or pulled ($[\varepsilon] \ne [0]$).   Any pushing/pulling direction is allowed; the elastic
response is taken into account by using expression (\ref{isotropic}) for the
strain tensor of layer 2.

The deformation of layer $2$ introduces a mismatch between the two honeycomb 
lattices.  For infinitesimally small deformations $\varepsilon$, the
crystallographically correct unit cell of the bilayer --- the ``least common
multiple'' of the unit cells of the two layers --- becomes infinitely large. 
For practically feasible descriptions of the asymetrically distorted bilayer one
is forced to choose deformations that lead to not-too-large unit cells,
resulting in a discrete set of strain values $\varepsilon$.  For example, for
their ab initio calculations, Choi {\it et al.} \cite{Cho} considered minimal
zig-zag strains of about $2\%$, requiring a supercell of $51$ unit cells in layer $1$ and $50$
in layer $2$.  A tight-binding description faces the same mismatching
problem.  Apart from the necessity to set up a larger hamiltonian matrix, the
inter-layer hopping parameters would have to be carefully reconsidered.  Indeed,
in the case of asymmetric strain, the A$_2$ atom originally directly above a
particular A$_1$ atom now has a different position, while a B$_1$ atom may now
have a direct (or almost direct) neighbor above.
In a way, the structure becomes a mix of AB and AA stacking \cite{Cho}.
A similar issue was addressed recently in Ref.\ \cite{San}: spatial modulations in the bilayer interlayer hopping arising due to elastic shearing or twisting were shown to lead to a non-Abelian gauge potential in the description of the low-energy electronic spectrum.

Designing the full
tight-binding matrix with correctly interpolated inter-layer hopping parameters
is  a formidable task, the outcome of which would still only be a limited
set of finite accessible strains $\varepsilon$.
As a compromise, we therefore consider the following small-$\vec{k}$ hamiltonian:

{\scriptsize
\begin{multline}
{\mathfrak H}(\vec{k}) =    \left(\begin{array}{cccc}
 V  & \hbar v_\text{F} k e^{i\phi}  & t_\perp & 0 \\
\hbar v_\text{F} k e^{-i\phi}   & V  & 0 & 0 \\
 t_\perp & 0 &
V + \frac{3a}{2}\frac{\partial V}{\partial a}(\varepsilon_{xx} + \varepsilon_{yy})
  &  \hbar v_\text{F} k e^{-i\phi} + i
\frac{3a}{4}\frac{\partial t}{\partial a} (-\varepsilon_{xx} + 
\varepsilon_{yy} -
2i\varepsilon_{xy}) \\
 0 & 0 & \hbar v_\text{F} k e^{i\phi} - i \frac{3a}{4}\frac{\partial t}{\partial a} (-\varepsilon_{xx} + 
\varepsilon_{yy} +
2i\varepsilon_{xy}) & V + \frac{3a}{2}\frac{\partial V}{\partial a}(\varepsilon_{xx} + \varepsilon_{yy})
  \\
  \end{array}\right), \\ \label{hamIIIa}
\end{multline}
}

\hspace*{-\parindent}where the parameter $t_\perp$ has to be interpreted as representing the average
hopping between the two layers.
The application of asymmetric plain affects both
the diagonal and the off-diagonal elements; the resulting band structure is
therefore non-trivial and worth investigating.  Diagonalising the hamiltonian
(\ref{hamIIIa}) results in the following secular equation:
\begin{align}
  \bigl(E(\vec{k})^2 - |\zeta(\vec{k})|^2\bigr)\Bigl[ \bigl(F -
E(\vec{k})\bigr)^2 - |\zeta(\vec{k}) + \delta \zeta|^2  \Bigr] + t_\perp^2
E(\vec{k}) \bigl(F - E(\vec{k})\bigr) = 0, \label{seceq1}
\end{align}
with
\begin{subequations}
\begin{align}
  F & = \frac{3a}{2}\frac{\partial V}{\partial a}(\varepsilon_{xx} + \varepsilon_{yy}) = 
      \frac{3a}{2}\frac{\partial V}{\partial a} \varepsilon (1 - \sigma ),  \label{seceq} \\
  \zeta(\vec{k}) & = \frac{3at}{2} k e^{i\phi}, \\
  \delta \zeta & = - i \frac{3a}{4}\frac{\partial t}{\partial a}
(-\varepsilon_{xx} + \varepsilon_{yy} + 2i\varepsilon_{xy}) = 
        \frac{3a}{4}\frac{\partial t}{\partial a} \varepsilon \bigl[i (1 -
\sigma)\cos(2\theta)  +  (1 + \sigma)\sin (2\theta)\bigr]. \label{seceq4}
\end{align}
\end{subequations}
Note that the argument $\phi = \tan^{-1}\frac{k_y}{k_x}$ does, in general, not
cancel out, and that the full 2D dependence of $E(k_x, k_y)$ has to be
considered.
Note also that upon replacing $\varepsilon$ by $-\varepsilon$ in Eqs.\
(\ref{seceq1}) -- (\ref{seceq4}),
the secular equation remains invariant if $E$ changes sign and the phase $\phi$
is shifted by $\pi$.  Hence, the substitution $\varepsilon \longrightarrow
-\varepsilon$ only leads to a band inversion and it suffices, as far as the
opening of a band gap concerns, to consider $\varepsilon \ge 0$.

The change in band structure comes from the interplay between the
values of $a \frac{\partial V}{\partial a}$, $a\frac{\partial t}{\partial
a}$ and $t_\perp$. To obtain an estimate for $\frac{\partial t}{\partial a}$ we
use the equivalent of relation (\ref{relchange}):
\begin{align}
  \frac{\partial t}{\partial a} = -2\frac{t}{a} = -4.394\text{ eV/{\AA}}. \label{relchange2}
\end{align}
Note that Harrison's relation [Eqs.\ (\ref{relchange}) and (\ref{relchange2})] should be taken as a rule of thumb rather than as an exact result \cite{Har}.  Other (experimental or theoretical) values than $2$ for $\eta = -\frac{\partial \ln t}{\partial \ln a}$ circulate in the literature (e.g.\ $\eta = 3.6$ \cite{Pie}, $\eta = 3$ \cite{Bas} or $\eta = 1.1$ \cite{Her}).  As it is our aim to make qualitative conclusions, we have used $\eta = 2$.

\begin{figure*}
\subfigure[]{
\resizebox{8cm}{!}{\includegraphics{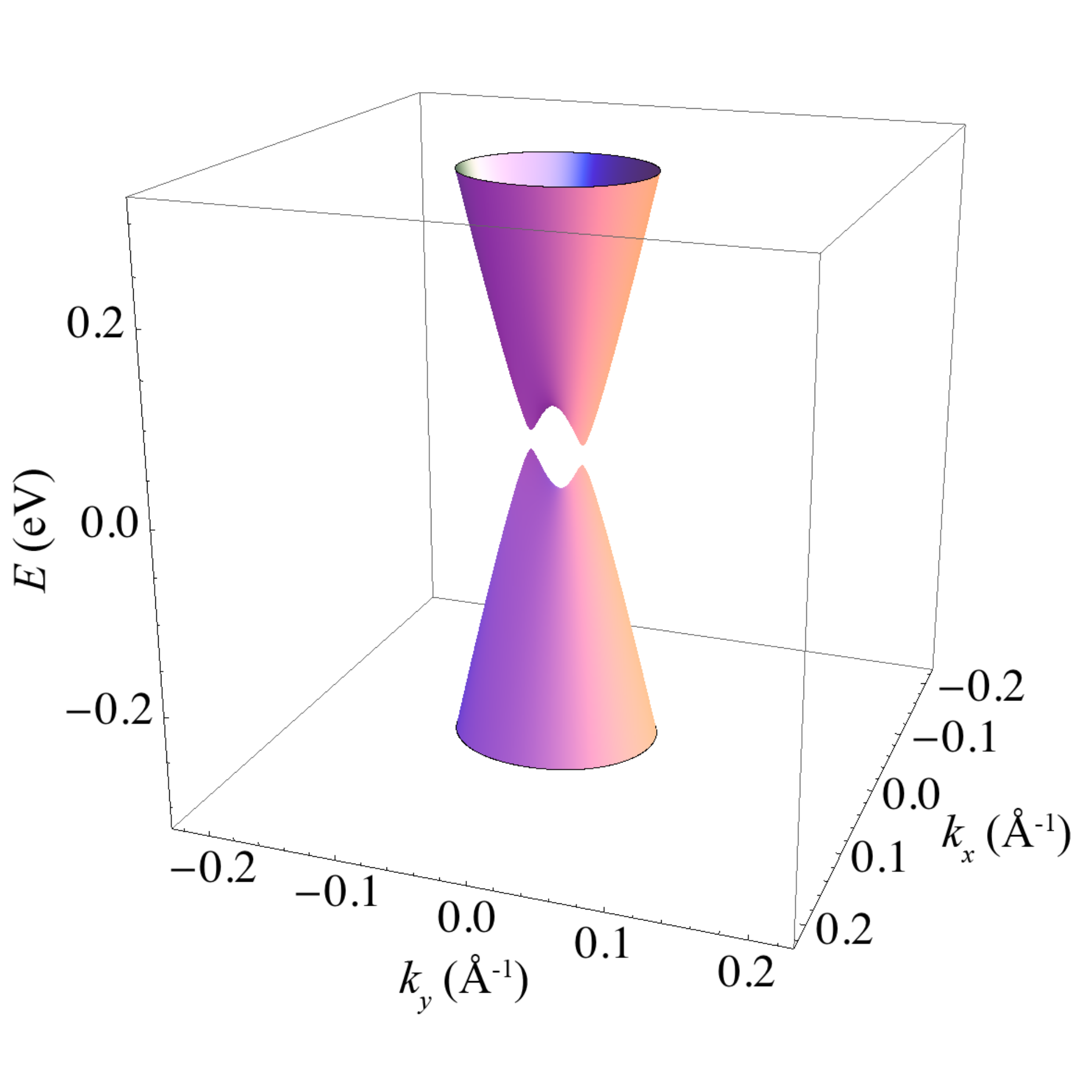}}}
\subfigure[]{
\resizebox{8cm}{!}{\includegraphics{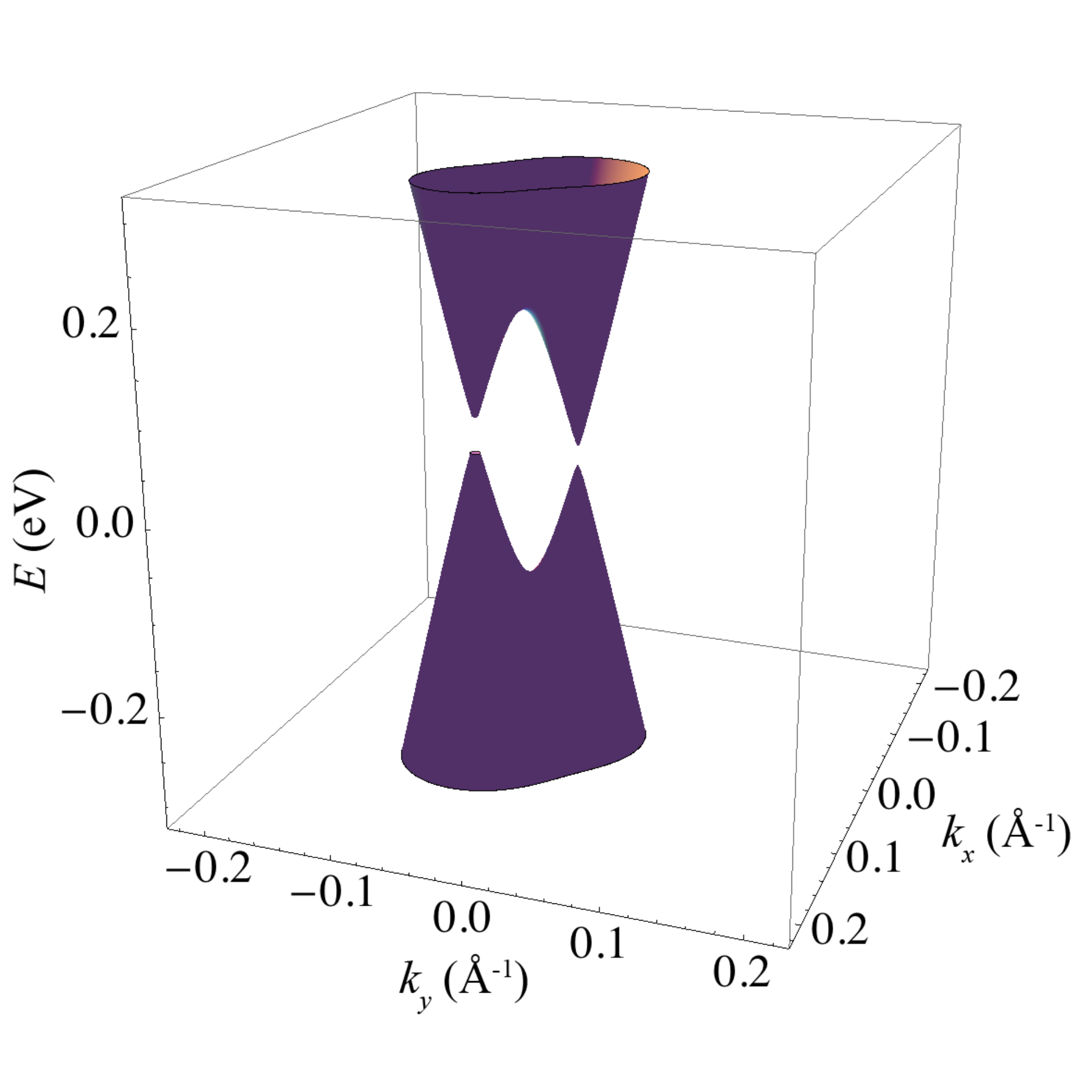}}}
\subfigure[]{
\resizebox{8cm}{!}{\includegraphics{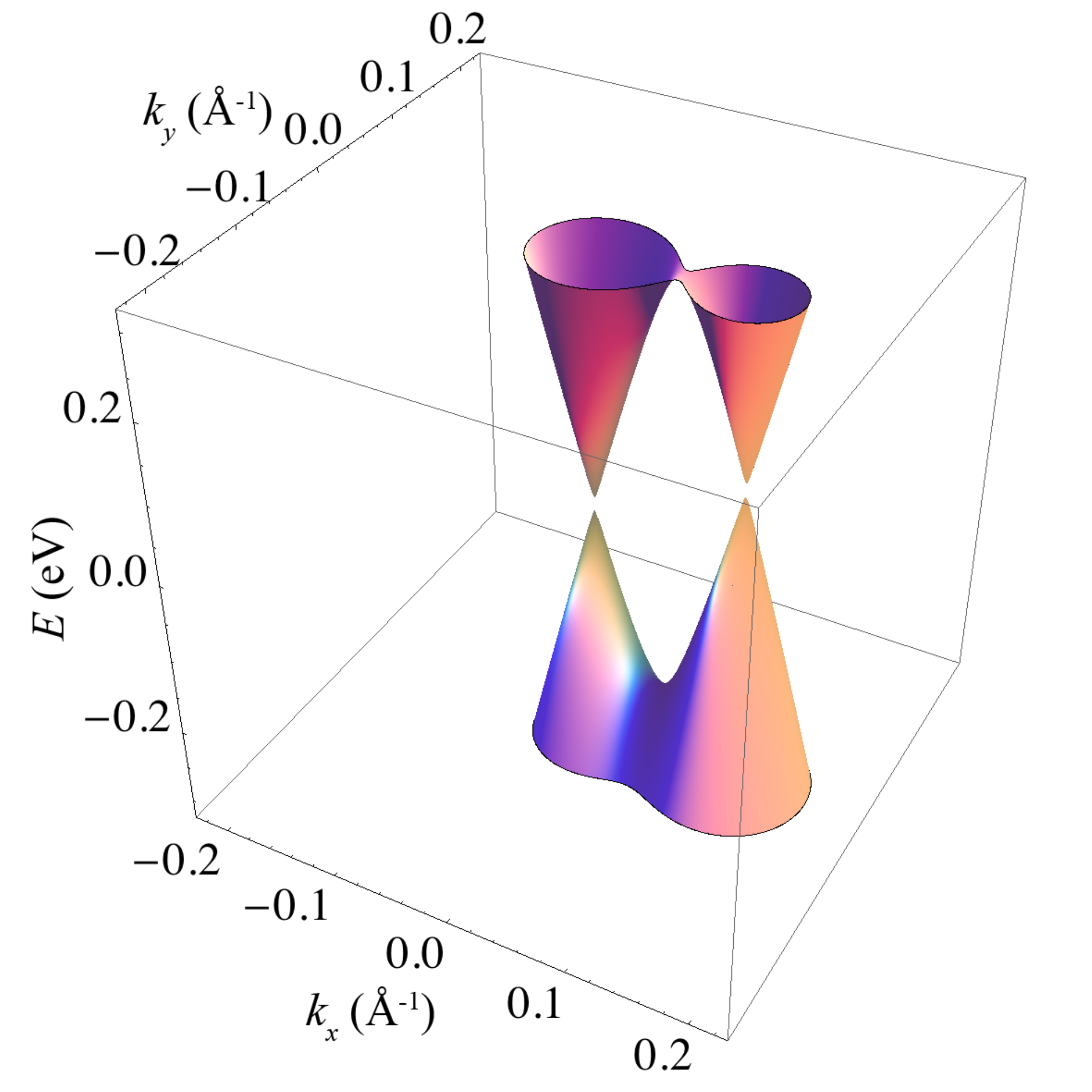}}}
\caption{Band structure of asymmetrically in-plane deformed bilayer graphene
near the K-point, for $\frac{\partial V}{\partial a} = 0.351$ eV/{\AA}: (a)
$\varepsilon = 0.05$, $\theta = 0$, (b) $\varepsilon = 0.1$, $\theta = \pi/8$,
(c) $\varepsilon = 0.15$, $\theta = \pi/4$.
\label{fig3d}}
\end{figure*}

For $\frac{\partial V}{\partial a}$, we proceed as for $\frac{\partial V}{\partial c}$ in Sect.\ \ref{sec3} and put a charge at $\vec{r} = \vec{0}$ and $\vec{r} = a' \vec{e}$, with $\vec{e}$ any unit vector perpendicular to the $z$-axis (a convenient choice is $\vec{e} = \vec{e}_x$) to mimic the periodic lattice potential :
\begin{align}
   U(\vec{r}) \propto -\frac{1}{|\vec{r}|} - \frac{1}{|\vec{r} - a'\vec{e}|} =
-\frac{1}{r} - \frac{1}{\sqrt{r^2 - 2a'r\sin\theta \cos\phi + a'^2}}. \label{propto2a}
\end{align}
Using the definition (\ref{definition}) for $V(a')$, we numerically calculate the
derivative $\left.\frac{\partial V}{\partial a}\right|_{a' = a}$ and obtain
$0.0439$ eV/{\AA} and $0.351$ eV/{\AA} for the cases  $V(a' = a) = -0.25$ eV and
$V(a' = a) = -2$ eV, respectively.
Inserting the value $\frac{\partial V}{\partial a} = 0.351$ eV/{\AA}, we
observe the opening of a band gap for arbitrarily small values of
$\varepsilon$. 
In Fig.\
\ref{fig3d}, low-$\vec{k}$ band structures are shown for a few choices of
$(\varepsilon,\theta)$.  The $(k_x,k_y)$ electronic dispersions always consist
of two facing doubly-peaked surfaces.  This can result in an indirect band
gap [Fig.\ \ref{fig3d}(a)], but also in indirect band crossings
[Figs.\ \ref{fig3d}(b) and (c)].  An inspection of the
ranges $0 \le \varepsilon \le 0.25$ and $0 \le \theta \le \pi$ allows to
conclude that (i) above a critical value for
the strain, only indirect band crossings are observed and (ii) the
angle $\theta$ has little or no influence on the presence/absence of a band gap.
The former conclusion agrees with the observation of Choi {\it et al.} \cite{Cho} of
a decrease of the band gap for zig-zag strains larger than $\varepsilon \approx
9\%$.
As for the dependence of the band gap on $\theta$, Choi {\it et al.}
\cite{Cho} found that in the case of armchair strains, no band gap appears at
all.  We recall that we allow for an elastic restoring force perpendicular to
the pulling/pushing direction which results in a dependence between
$\varepsilon_{xx}$, $\varepsilon_{xy}$ and $\varepsilon_{yy}$
[Eq.\ (\ref{isotropic})].  In our model, this isotropy makes the band gap
development independent of $\theta$.

\begin{figure*}
\resizebox{8cm}{!}{\includegraphics{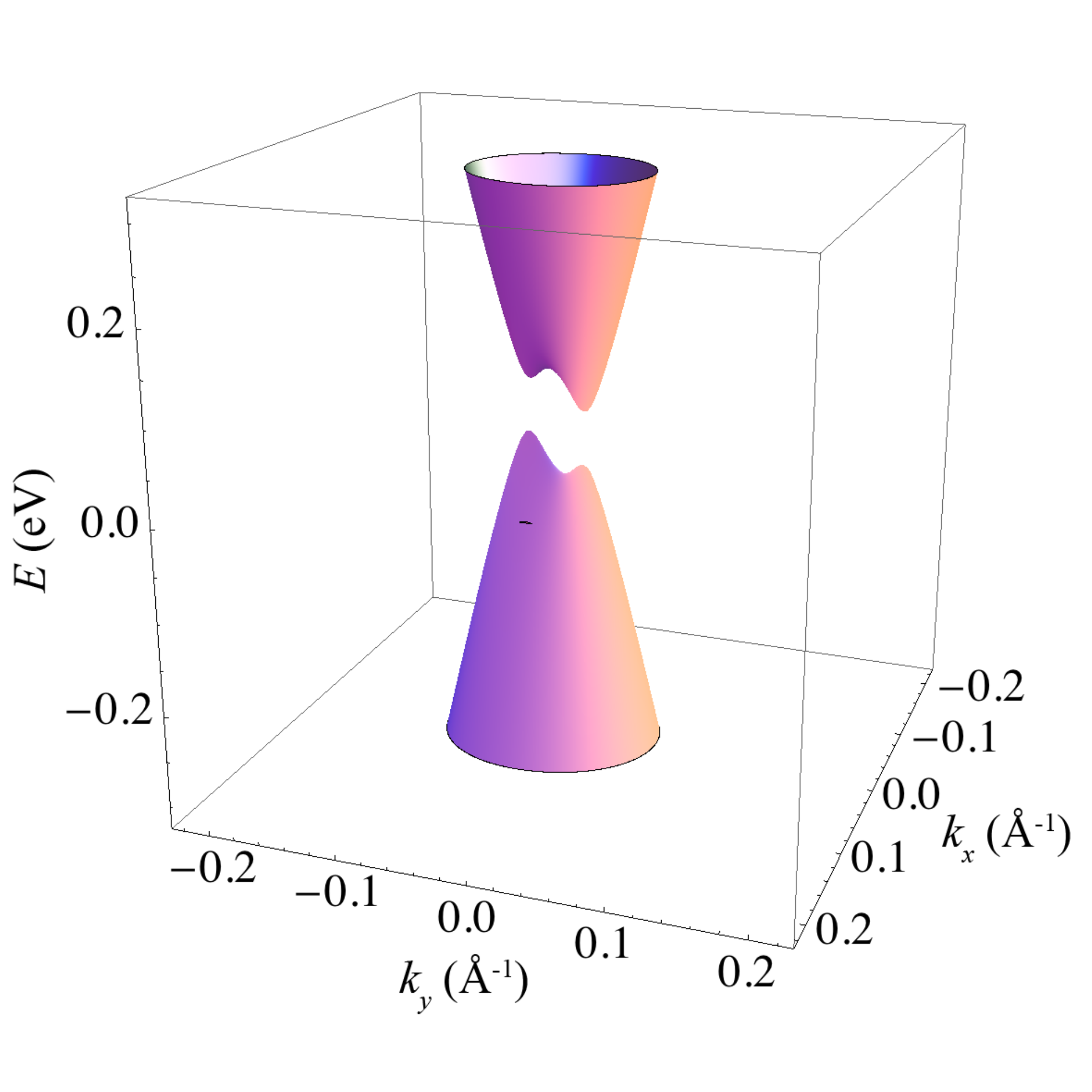}}
\caption{
Band structure of asymetrically in-plane deformed bilayer graphene
near the K-point, for $\frac{\partial V}{\partial a} = 1$ eV/{\AA}, 
$\varepsilon = 0.05$ and $\theta = 0$.
\label{fig3d2}}
\end{figure*}

A true (but indirect) band gap is only observed for small strains
$\varepsilon$, as e.g.\ in Fig.\ \ref{fig3d}(a) ($\varepsilon = 0.05$, $\theta
=0 $, $\Delta E_\text{g} \approx 10$ meV).  The maximal band gap turns out to
depend on the magnitude of $\frac{\partial V}{\partial a}$, not unlike the
situation in Sect.\ \ref{sec3} for transversal strain where the value of
$\frac{\partial V}{\partial c}$ is critical.  For
the value $\frac{\partial V}{\partial a} = 0.0439$ eV/{\AA}, corresponding to
$V(a) = -0.25$ eV, the band gap for $\varepsilon = 0.05$ and $\theta = 0
$ reduces to $\Delta E_\text{g} \approx 1$ meV).  For the artificially large
value of $\frac{\partial V}{\partial a} = 1$ eV, we find (Fig.\ \ref{fig3d2})
$\Delta E_\text{g} = 25$ meV.
At this point we remark that one can identify $\frac{3a}{2}\frac{\partial V}{\partial a}$ with the deformation potential $g_1$ entering the strained monolayer hamiltonian (\ref{hammono}), whose value for graphite is about $16$ eV \cite{Sug2}.  For
$\frac{\partial V}{\partial a}$ one therefore has approximately $7.5$ eV {\AA}, which suggests that larger values of 
$\frac{\partial V}{\partial a}$ may indeed be more appropriate.  Obviously, experiments and/or ab initio calculations should be carried out to clarify the parameters' values.

In conclusion, hamiltionian (\ref{hamIIIa}), a possible model for
describing the effect of asymmetric lateral strains in bilayer graphene, results
in indirect band gaps for small strains ($\varepsilon \approx 5 \%$).  For
larger strains ($\varepsilon \gtrsim 10\%$), overlapping indirect bands result
in a gapless spectrum.  No critical directional strain dependence is observed,
which comes from the anisotropy of the 2D hexagonal lattice.


\section{conclusions}
Applying strain to a graphene bilayer alters its electronic structure.  Using the simplest tight-binding model, featuring only nearest-neighbor intra- and inter-plane hopping ($t$ and $t_\perp$), we have shown that for the AB (Bernal) stacking of two graphene sheets, several band gap scenarios are possible.

A first possibility is to push (pull) the bilayer's two graphene sheets towards (away from) each other.  As a consequence of the different neighborhoods experienced by A and B carbon atoms, the change in on-site tight-binding energies results in a symmetry breaking which, for large enough strains, opens a band gap.  We find that pulling and pushing are inequivalent; the former is more effective in producing a gap.  Predictions of the minimally required strain to open a band gap critically depend on values for the changes in on-site and inter-layer hopping energies with inter-plane distance ($\frac{\partial V}{\partial c}$ and $\frac{\partial t_\perp}{\partial c}$).  These quantities are not readily available from the literature.  We therefore rely on rules of thumb to make acceptable estimates, and find that for pulling, a band gap opens from $\varepsilon_{zz} = 0.25$ onwards.  For $\varepsilon_{zz}\approx 30\%$, the band gap is about $125$ meV.
For pushing, the critical strain is well beyond $50\%$.  Note that a strain of $50\%$ would require a pressure of $p = 0.5c_{44} \approx 2$ GPa (taking for the elastic constant $c_{44}$ the value of graphite, $c_{44} = 4.18$ GPa \cite{MicVerPRB2008}), which is experimentally feasible.  While bringing the two layers close together can be realized by applying high pressure, pulling the graphene sheets away from each other is an experimental challenge.  A possible indirect way to do so would be to intercalate the bilayer; recently, for example, Li atoms have successfully been intercalated between two graphene sheets \cite{Sug}.
Interestingly, we find that at the critical strain, the electronic dispersion near the K-point consists of a triple degeneracy of two crossing linear bands (Dirac cone) and one parabolic band.  This result shows that under certain symmetry-breaking strain conditions, bilayer graphene can exhibit Dirac fermions.  Normally, Dirac fermions only occur in graphene multilayers consisting of an odd number of graphene sheets \cite{Par2}.  The asymmetry of the hamiltonian matrix diagonal lies at the basis of the band structure modifications observed in Fig.\ \ref{fig2}.  It was already 
recognized before \cite{Muc3} that intra-layer on-site energy differences can lead to band gaps of a different type than the
``Mexican-hat-like" band gap band structures \cite{McC2,Gui,Oht,McC3,Cas} associated with inter-layer energy differences e.g.\ induced by the application of a bias.  In the present work, we have shown that uniform perpendicular strain provides a physical means for enhancing the intra-layer energy differences.

A second possibility is to push/pull parallel to the bilayer (symmetric uniform strain).  It turns out that the criterion for opening a band gap is the same as for monolayer graphene; the inter-plane coupling $t_\perp$ does not play any role.  The conclusions made by Pereira {\it et al.} \cite{Per} for the monolayer can be transferred to the bilayer; a strain along the zig-zag direction larger than $\sim 23\%$ results in a band gap, while pushing/pulling along the armchair direction never leads to a gap.

Finally, we considered asymmetric uniform strain: the two graphene sheets experience different strains parallel to the bilayer.  Assuming a tight-binding hamiltonian matrix where $t_\perp$ represents an average inter-plane hopping, we obtain the interesting result that a small finite strain, in any direction, immediately opens a band gap.  The maximal band gap depends, similarly to the case of transversal strain, on the interplay of the quantities $\frac{\partial V}{\partial a}$ and $\frac{\partial t}{\partial a}$, reliable estimates for which are hard to derive.
The observed band gaps are relatively small --- $\Delta E_\text{g} \approx 10$ meV.  The advantage, however, is that there is no strain barrier.  Rather, beyond a certain strain (typically $10\%$), bands start to indirectly overlap and destroy the gap.
As opposed to the case of uniform symmetric strain, no dependence on the strain direction was obtained --- a consequence of the isotropy of a hexagonal lattice and of taking the elastic response in the perpendicular direction into account.  Pushing and pulling are equivalent.  We point out that the observed band gaps are indirect, which is relevant for possible opto-electronic applications.  Our results are in qualitative agreement with recent ab initio calculations of the electronic structure of similarly asymmetrically strained bilayer graphene \cite{Cho}.

Although having provided estimates for strains $\varepsilon$ required for the opening of a band gap and associated band gap magnitudes $\Delta E_\text{g}$, we wish to emphasize the qualitative aspect of our results.  Irrespective of the precise values of the tight-binding parameters and derived quantities, it is the particular structure of the graphene bilayer that, when deformed, allows for gapped electronic structures.  In our opinion, the various ways shown here in which a band gap can be induced in bilayer graphene by deformations should stimulate experimental investigations on the possibility of strain-engineering bilayer graphene's electronic properties.  Recalling graphene's high strength, this should be feasible.  In addition, we suggest that the theoretical models presented here be reconsidered by means of precise ab initio calculations.

\acknowledgments
The authors would like to acknowledge O. Leenaerts, E. Mariani, K.H. Michel and J. Schelter for useful discussions.  B.V. was financially supported by the Flemish Science Foundation (FWO-Vl).  This work was financially supported by the ESF programme EuroGraphene under projects CONGRAN and ENTS as well as by the DFG.

\appendix
\section{Strain}\label{appA}
As described in the main text, deformations of the bilayer graphene carbon
network enter the tight-binding formalism via bond length changes [Eqs.\
(\ref{bc1}) -- (\ref{bc4})]. 
In this Appendix, we elaborate the expressions for $\delta a_l$ and $\delta c$
and
the strained graphene bilayer tight-binding hamiltonian.

\subsection{Continuous description.  Long-wavelength limit}
In general, one has
\begin{align}
  a + \delta a_l & = \Bigl|\vec{X} + \vec{d}_l + \vec{u}(\vec{X} +
\vec{d}_l)
 - \bigl(\vec{X} + \vec{u}(\vec{X})\bigr)\Bigr| \nonumber \\
 & = \sqrt{a^2 + 2 \vec{d}_l\cdot \bigl( \vec{u}(\vec{X} +
\vec{d}_l)
 -  \vec{u}(\vec{X}) \bigr) + \bigl| \vec{u}(\vec{X} +
\vec{d}_l )
 -  \vec{u}(\vec{X})\bigr|^2}.
\end{align}
Assuming {\em small bond length changes} $\delta a_l$, we retain only
first-order terms:
\begin{align}
 a + \delta a_l \approx a \left(1 + \frac{1}{a^2} \vec{d}_l\cdot \bigl(
\vec{u}(\vec{X} +
\vec{d}_l)  -  \vec{u}(\vec{X}) \bigr)
 \right), \label{A2}
\end{align}
so that
\begin{align}
 \delta a_l \approx \frac{1}{a} \vec{d}_l\cdot \bigl(
\vec{u}(\vec{X} +
\vec{d}_l)  -  \vec{u}(\vec{X}) \bigr)
.
\end{align}

In the case of small bond length changes, the differences
between displacements at neighboring sites $\bigl|\vec{u}(\vec{X} +
\vec{d}_l) - \vec{u}(\vec{X})\bigr|$ must be small as well.  In other
words, the
displacement field $\vec{u}$ is a slowly varying (vector) function of $\vec{X}$,
implying
that a
linearisation of $\vec{u}$ at $\vec{X}$ approximates $\vec{u}$ at
$\vec{X} + \vec{d}_l$ well enough.  In this so-called {\em
long-wavelength limit} we then have
\begin{align}
 \vec{d}_l \cdot \bigl( \vec{u} (\vec{X} + \vec{d}_l) -
\vec{u} (\vec{X})\bigr)
  & \approx  \left. (\vec{d}_l\cdot
\vec{\nabla}) (\vec{d}_l \cdot \vec{u}  )\right|_{\vec{X}}.
\end{align}
The operator $\vec{\nabla} = \left(\frac{\partial}{\partial x},
\frac{\partial}{\partial y}, \frac{\partial}{\partial z}\right)$ acts on
$\vec{u}(\vec{X})$ with $\vec{X}$ taken as the continuous spatial variable
$\vec{x} = (x,y,z)$.
Writing $d_{l\alpha}$, $\alpha = 1, 2$ for the $x$- and $y$-components of
$\vec{d}_l$ ($\vec{d}_l$ has no $z$-component, see top of Fig.\ \ref{fig1}),
we obtain
\begin{align}
   \left. (\vec{d}_l\cdot
\vec{\nabla}) (\vec{d}_l \cdot \vec{u}  )\right|_{\vec{X}} & =
\sum_{\alpha = 1}^2 \sum_{\beta = 1}^2
d_{l\alpha}d_{l\beta}\left.\frac{\partial u_\beta}{\partial
x_\alpha}\right|_{\vec{X}}.
\end{align}
In the following, we drop the $\left.\right|_{\vec{X}}$ attributes and
keep in mind that there formally is a dependence on the position in the
lattice.  The expression for $\delta a_l$ now becomes
\begin{align}
 \delta a_l \approx \frac{1}{a} \sum_{\alpha = 1}^2 \sum_{\beta = 1}^2
d_{l\alpha}d_{l\beta}
\frac{\partial u_\beta}{\partial
x_\alpha}.
\end{align}
Calculating explicit expressions for the quantities 
$D_{\alpha\beta}^{l} = d_{l\alpha} d_{l\beta}$ leads to
\begin{xalignat}{3}
   D^1  = \frac{a^2}{4}\left( \begin{array}{cc} 1 & \sqrt{3} \\ \sqrt{3} & 3 \\
\end{array}\right), && D^2 = \frac{a^2}{4}\left( \begin{array}{cc} 1 &
-\sqrt{3}
\\ -\sqrt{3} & 3 \\
\end{array}\right), && D^3 = a^2\left( \begin{array}{cc} 1 & 0
\\ 0 & 0 \\
\end{array}\right).
\end{xalignat}

In a completely analogous way we obtain for $\delta c$ --- recalling that
$\vec{c} = c\vec{e}_z$ --- the following expression (up to first order):
\begin{align}
 \delta c =  \Bigl|\vec{X} + \vec{c}  + \vec{u}(\vec{X} +
\vec{c})
 - \bigl(\vec{X} + \vec{u}(\vec{X})\bigr)\Bigr| - c
 \approx  
c 
\frac{\partial u_z}{\partial
z}
.
\end{align}

\subsection{Corrections to the hamiltonian matrix}

The corrections to the hamiltonian involve the following quantities:
\begin{subequations}
\begin{align}
  \delta U_\text{A} & =  \sum_{l = 1}^3 \frac{\partial V}{\partial
a }  \delta a_l +  \frac{\partial V}{\partial c} \delta c =
\frac{3a}{2}\frac{\partial V}{\partial
a }(\varepsilon_{xx} + \varepsilon_{yy}) + c \frac{\partial V}{\partial c}
\varepsilon_{zz},
 \\
 \delta U_\text{B} & = \sum_{l = 1}^3 \frac{\partial V}{\partial
a }  \delta a_l = \frac{3a}{2}\frac{\partial V}{\partial
a }(\varepsilon_{xx} + \varepsilon_{yy}).
\end{align}
\end{subequations}
Here, use has been made of the identity
\begin{align}
 \sum_{l = 1}^3 D^{l} = \frac{3a^2}{2}\left( \begin{array}{cc} 1 & 0 \\
0 & 1 \\
\end{array}\right)
\end{align}
and the definition of the (first-order) components of the strain tensor:
\begin{align}
  \varepsilon_{ij} = \frac{1}{2}\left(\frac{\partial u_i}{\partial x_j} + 
\frac{\partial u_j}{\partial x_i}
\right). \label{straintensor2}
\end{align}
Note that the strain tensor is, in general, space-dependent: $[\varepsilon]
\equiv [\varepsilon](\vec{X})$.  Therefore, transforming the summation
\begin{align}
  \sum_{\vec{X}_1}c_{\vec{X}_1}^\dagger \delta U^\text{A}c_{\vec{X}_1} = 
 \sum_{\vec{X}_1}c_{\vec{X}_1}^\dagger \left[ \frac{3a}{2}\frac{\partial
U}{\partial
a }(\varepsilon_{xx} + \varepsilon_{yy}) + c \frac{\partial V}{\partial c}
\varepsilon_{zz}\right]c_{\vec{X}_1}
\end{align}
into a summation over $\vec{q}$-vectors involves discrete Fourier transforms:
\begin{subequations}
\begin{align}
  \sum_{\vec{X}_1}c_{\vec{X}_1}^\dagger \delta U^\text{A}c_{\vec{X}_1} & = 
 \frac{1}{\sqrt{N}}\sum_{\vec{q}}\sum_{\vec{q}'}
c_{\text{A}_1}^\dagger(\vec{q})\left[
\frac{3a}{2}\frac{\partial
U}{\partial
a }\bigl(\varepsilon_{xx}(\vec{q}' - \vec{q}) + \varepsilon_{yy}(\vec{q}' -
\vec{q})\bigr) + c \frac{\partial V}{\partial c}
\varepsilon_{zz}(\vec{q}' - \vec{q})
\right]
c_{\text{A}_1}(\vec{q}'), \label{Ac1} \\
\varepsilon_{ij}(\vec{q})& =
\frac{1}{\sqrt{N}}\sum_{\vec{X}_i}\varepsilon_{ij}(\vec{X}_i)e^{i \vec{q} \cdot
\vec{X}_i}. \label{Ac2}
\end{align}
\end{subequations}
Importantly, when the strain tensor $[\varepsilon]$ is space-independent, i.e.\ in the case of
uniform deformations, Eqs.\ (\ref{Ac1}) and (\ref{Ac2}) collapse into
\begin{align}
  \sum_{\vec{X}_1}c_{\vec{X}_1}^\dagger \delta U^\text{A}c_{\vec{X}_1} & = 
 \sum_{\vec{q}} c_{\text{A}_1}^\dagger(\vec{q})\left[
\frac{3a}{2}\frac{\partial
U}{\partial
a }(\varepsilon_{xx}  + \varepsilon_{yy} ) + c \frac{\partial
U}{\partial c}
\varepsilon_{zz}
\right]
c_{\text{A}_1}(\vec{q})
\end{align}
so that
\begin{align}
 \delta {\mathfrak H}_{11} = \frac{3a}{2}\frac{\partial
U}{\partial
a }(\varepsilon_{xx}  + \varepsilon_{yy} ) + c \frac{\partial
U}{\partial c}
\varepsilon_{zz}
\end{align}
is $\vec{q}$-independent.  For the correction to ${\mathfrak H}_{22}$
we then have
\begin{align}
  \delta {\mathfrak H}_{22} = \frac{3a}{2}\frac{\partial
U}{\partial
a }(\varepsilon_{xx}  + \varepsilon_{yy}).
\end{align}

The off-diagonal non-zero hopping matrix elements are
more complicated. We will need the quantities
\begin{subequations}
\begin{align}
 \delta t_l & =  \frac{\partial t}{\partial
a} \delta a_l =  \frac{1}{a} \frac{\partial t}{\partial
a} \sum_{\alpha = 1}^2\sum_{\beta = 1}^2
D^l_{\alpha\beta}\frac{\partial u_\beta}{\partial x_\alpha}, \\
  \delta t_\perp & =  \frac{\partial t_\perp}{\partial
c} \delta c = c \frac{\partial t_\perp}{\partial
c} 
\frac{\partial u_z}{\partial
z}. 
\end{align}
\end{subequations}

Let us consider the term
\begin{align}
-  \sum_{\vec{X}_1}\sum_{l = 1}^3 c_{\vec{X}_1}^\dagger \delta t_l c_{\vec{X}_1
+ \vec{d}_l} & =
-\frac{1}{\sqrt{N}}\frac{1}{a} \frac{\partial t}{\partial
a}\sum_{\vec{q}}\sum_{\vec{q}'}\sum_{\alpha = 1}^2\sum_{\beta =
1}^2
c_{\text{A}_1}^\dagger(\vec{q})
e_{\alpha\beta}(\vec{q}' - \vec{q})
F_{\alpha\beta}(\vec{q}')
c_{\text{B}_1}(\vec{q}'), \label{foregoing}
\end{align}
with
\begin{subequations}
\begin{align}
  F_{\alpha\beta}(\vec{q}) & = \sum_{l = 1}^3 D^l_{\alpha
\beta}e^{i\vec{q} \cdot \vec{d}_l}, \\
 e_{\alpha\beta}(\vec{q}) & = \frac{1}{\sqrt{N}}\sum_{\vec{X}_1}e^{i
\vec{q} \cdot \vec{X}_1} \frac{\partial u_\beta}{\partial x_\alpha}.
\end{align}
\end{subequations}
Again, in the case of uniform deformations, $\frac{\partial u_\beta}{\partial
x_\alpha}$ is space-independent and Eq.\ (\ref{foregoing}) simplifies to
\begin{align}
-  \sum_{\vec{X}_1}\sum_{l = 1}^3 c_{\vec{X}_1}^\dagger \delta t_l c_{\vec{X}_1
+ \vec{d}_l} & =
- \frac{1}{a} \frac{\partial t}{\partial
a}\sum_{\vec{q}} c_{\text{A}_1}^\dagger(\vec{q}) \sum_{\alpha = 1}^2\sum_{\beta
=
1}^2
F_{\alpha\beta}(\vec{q}) \frac{\partial u_\beta}{\partial
x_\alpha}
c_{\text{B}_1}(\vec{q}),
\end{align}
so that the matrix element ${\delta \mathfrak H}_{12}$ becomes
\begin{align}
  \delta {\mathfrak H}_{12} = - \frac{1}{a} \frac{\partial t}{\partial
a}  \sum_{\alpha = 1}^2\sum_{\beta =
1}^2
F_{\alpha\beta}(\vec{q}) \frac{\partial u_\beta}{\partial
x_\alpha}.
\end{align}
For $F(\vec{q})$ one obtains
\begin{align}
  F(\vec{q}) & = a^2\left(\begin{array}{cc}
  \frac{1}{4}( e^{i\vec{q}\cdot \vec{d}_1} +  e^{i\vec{q}\cdot \vec{d}_2}) +  e^{i\vec{q}\cdot \vec{d}_3}& 
   \frac{\sqrt{3}}{4} (e^{i\vec{q}\cdot \vec{d}_1} -  e^{i\vec{q}\cdot \vec{d}_2}) \nonumber \\
   \frac{\sqrt{3}}{4} (e^{i\vec{q}\cdot \vec{d}_1} -  e^{i\vec{q}\cdot \vec{d}_2}) &  \frac{3}{4} (e^{i\vec{q}\cdot \vec{d}_1} +  e^{i\vec{q}\cdot \vec{d}_2}) \\
   \end{array}\right) \\
   & = a^2 e^{i \frac{1}{2}a q_x}  \left(\begin{array}{cc}
  \frac{1}{2}   \cos (\frac{\sqrt{3}}{2}a q_y)  + e^{-i  \frac{3}{2}a q_x} & 
   \frac{\sqrt{3}}{2} i   \sin (\frac{\sqrt{3}}{2}a q_y) \\
  \frac{\sqrt{3}}{2} i  \sin (\frac{\sqrt{3}}{2}a q_y) & \frac{3}{2}   \cos (\frac{\sqrt{3}}{2}a q_y) \\
   \end{array}\right).
\end{align}
Near the K-point,
 one has
\begin{align}
 F_{\alpha\beta}(\vec{q}_\text{K} + \vec{k}) = \frac{3a^2}{4}e^{i \frac{\pi}{3}}
\left(\begin{array}{cc}
-1 + \frac{a}{2}(3i k_x - k_y) & i - \frac{a}{2}(k_x - i k_y) \\ i - \frac{a}{2}(k_x - i k_y) & 1 + \frac{a}{2}(ik_x - 3k_y) \end{array}\right).
\end{align}
The phase factor $e^{i \frac{\pi}{3}}$ can be eliminated by redefining the
creation and annihilation operators for electrons at B sites (position vectors
$\vec{X}_1 + \vec{d}_l$ and $\vec{X}_2 - \vec{d}_l$):
\begin{subequations}
\begin{align}
 c_{\vec{X}_i \pm \vec{d}_l}^\dagger & \longrightarrow
 c_{\vec{X}_i \pm \vec{d}_l}^\dagger e^{- i\frac{\pi}{6}}, \label{redef1}
\\
 c_{\vec{X}_i \pm \vec{d}_l} & \longrightarrow 
c_{\vec{X}_i \pm \vec{d}_l}
e^{i\frac{\pi}{6}}. \label{redef2}
\end{align}
\end{subequations}
Near the K-point, we obtain the following correction to the matrix element:
\begin{align}
  \delta {\mathfrak H}_{12}(\vec{k}) = -i \frac{\partial t}{\partial
a} \frac{3a}{4}
 \Bigl(\varepsilon_{xx} \bigl[ -1 + \frac{a}{2}(3i k_x - k_y) \bigr]  + \varepsilon_{yy} \bigl[1 + \frac{a}{2}(ik_x - 3k_y)  \bigr]  + 2 \varepsilon_{xy} \bigl[ i - \frac{a}{2}(k_x - i k_y) \bigr]
  \Bigr),
\end{align}
where the factor $i$ comes from multiplying the phase factors
$e^{i \frac{\pi}{3}}$ and $e^{i \frac{\pi}{6}}$, and where the definition of
the strain tensor in Eq.\ (\ref{straintensor2}) has been used.
For small deformations,
the terms proportional to $\varepsilon_{\alpha\beta} k$, with $\alpha,\beta \in \{x,y\}$, can be neglected:
\begin{align}
  \delta {\mathfrak H}_{12}^\text{K} = -i \frac{\partial t}{\partial
a} \frac{3a}{4}
 (-\varepsilon_{xx} + \varepsilon_{yy} + 2i\varepsilon_{xy}).
\end{align}

We now consider the term
\begin{subequations}
\begin{align}
  \sum_{\vec{X}_1} \delta t_\perp  c_{\vec{X}_1}^\dagger    c_{\vec{X}_1
+ \vec{c}} 
 & =
\frac{1}{\sqrt{N}} c \frac{\partial t_\perp}{\partial
c} \sum_{\vec{q}} \sum_{\vec{q}'}
 c_{\text{A}_1}^\dagger(\vec{q}) f(\vec{q}' - \vec{q})  e^{i \vec{q}' \cdot
\vec{c}}
 c_{\text{A}_2} (\vec{q}'),
 \\
 f(\vec{q}) & = \frac{1}{\sqrt{N}} \sum_{\vec{X}_1} e^{i \vec{q}\cdot
\vec{X}_1} \frac{\partial u_z}{\partial
z}.
\end{align}
\end{subequations}
Here, the factor $e^{i \vec{q}' \cdot \vec{c}}$ is equal to $1$ because
$\vec{q}'$ (only $x$- and $y$-components) and $\vec{c}$ (only $z$-component)
are perpendicular.  For space-independent $\frac{\partial u_z}{\partial
z}$ we obtain
\begin{align}
  \sum_{\vec{X}_1} \delta t_\perp    c_{\vec{X}_1}^\dagger c_{\vec{X}_1
+ \vec{c}} & = c \frac{\partial t_\perp}{\partial
c} \frac{\partial u_z}{\partial
z} \sum_{\vec{q}}  
 c_{\text{A}_1}^\dagger(\vec{q}) 
 c_{\text{A}_2} (\vec{q}) = c \frac{\partial t_\perp}{\partial
c} \varepsilon_{zz} \sum_{\vec{q}}  
 c_{\text{A}_1}^\dagger(\vec{q}) 
 c_{\text{A}_2} (\vec{q}),
\end{align}
so that the correction to the matrix element ${\mathfrak H}_{13}$ reads
\begin{align}
 \delta {\mathfrak H}_{13} = c \frac{\partial t_\perp}{\partial
c} \varepsilon_{zz}.
\end{align}

These considerations then lead to the correction $\delta {\mathfrak H}$
to the
hamiltionan matrix ${\mathfrak H}$ given by Eqs.\ (\ref{corr1}) -- (\ref{corr1with}).

\end{document}